\documentstyle[aps,prl]{revtex}
\begin{document}
\twocolumn[\hsize\textwidth\columnwidth\hsize\csname @twocolumnfalse\endcsname
\title{Anomalous Conductance Distribution in Quasi-One Dimension: Possible
Violation of One-Parameter Scaling Hypothesis}
\author{Pritiraj Mohanty$^1$ and Richard A. Webb$^2$ \\
{\it $^1$ Department of Physics, Boston University,
590 Commonwealth Avenue, Boston, MA 02215 \\
$^2$ Department of Physics, Center for Superconductivity Research, University
of Maryland, College Park, MD 20742}}

\maketitle

\begin{abstract}
We report measurements of conductance distribution in a set of
quasi-one-dimensional gold wires. The distribution includes the 
second cumulant or the variance which describes the universal 
conductance fluctuations, and the third cumulant which denotes 
the leading deviation. We have observed an asymmetric 
contribution---or, a nonvanishing third cumulant---contrary to 
the expectation for quasi-one-dimensional systems in the 
noninteracting theories in the one-parameter scaling framework, 
which include the perturbative diagrammatic calculations and 
the random matrix theory.

\vskip 0.1in
\end{abstract}
]
 \makeatletter
\global\@specialpagefalse
\def\@oddhead{{\hskip 5in {\it Phys. Rev. Lett.} {\bf 88}, 146601 (2002)}\hfill}
\let\@evenhead\@oddhead
\makeatother      

\par

Scaling theory of Anderson localization is the starting point of
mesoscopic physics \cite{anderson1}. It forms the basis of our 
understanding of low-dimensional metals as well as various 
conductor-insulator transitions. The theory is based on the 
one-parameter scaling hypothesis which argues that the 
conductance $G$ is the only relevant parameter that controls 
its variation with size $L$, implying
\begin{equation}
{\partial \ln g \over \partial \ln L} = \beta(g); \quad\quad g \equiv {G \over e^2/h},
\end{equation}
\noindent
where $\beta(g)$ is a universal function of the dimensionless 
conductance $g$ alone. The perturbative calculations \cite{larkin} 
in terms of $1/g$ find the following Gell-Mann-Low $\beta(g)$-function for small 
deviations from the ohmic behavior in good metals ($g \gg 1$):
\begin{equation}
\beta(g) = (d-2) + \alpha/g + \cdots
\end{equation}
The first term represents the Ohm's law, and the second one is the 
leading quantum correction---for example, from weak localization 
($\alpha \simeq -1$). The scaling hypothesis 
has been confirmed by field-theoretical calculations 
in the nonlinear $\sigma$-model \cite{wegner}. 

One-parameter scaling must be understood in terms of the entire 
conductance distribution; this was realized almost immediately after its 
discovery \cite{anderson2}. Another reason for
studying conductance distribution for the interpretation of one-parameter scaling 
pertains to the sample-specific conductance fluctuations \cite{webb1,lee1}
with variance $\langle \delta g^2 \rangle \sim 1$, the discovery of 
which raised the problem of how to sum up a diverging series of quantum 
corrections of order one ($\delta G \sim e^2/h$) 
in low dimensions $d \le 2$. However, the validity of the perturbation 
theory (in $1/g$) was established, contingent upon one-parameter 
scaling \cite{akl}. Large conductance fluctuations were found to cause
an instability of the one-parameter scaling near 
the localization transition for $g \le 1$, leading to fluctuations
which deviate from gaussian to one with a log-normal tail. However, in the
metallic region ($g \gg 1$), for small fluctuations the
distribution was found to be gaussian \cite{akl}. These 
fluctuations are essentially the universal conductance fluctuations (UCF),
independent of sample parameters such as mean free path $l_e$
and average conductance $\langle g \rangle$ (not including the
effects  of temperature and dephasing). Further support for
one-parameter scaling in presence of conductance fluctuations
was obtained by calculations in random matrix 
theory \cite{stone,ohtsuki}. In general, in the metallic regime,
\begin{equation}
\langle g^n \rangle \propto \langle g \rangle^{2-n}, \quad n < g_0.
\end{equation}    
$g_0$ is the mean conductance at the scale $l_e$. For $n=2$, 
one obtains the UCF, independent of $\langle g \rangle$. Higher cumulants 
for $n > 2$ are small, and the distribution is dominantly gaussian.
For $n > g_0$, the magnitudes of higher order cumulants increase
rapidly as $\langle g^n \rangle \propto e^{n^2/g_0}$; this
leads to the log-normal distribution \cite{muttalib} in the non-metallic region 
near localization ($g \le 1$).

Any deviation from the normal (gaussian) distribution can be divided 
into two characteristic parts: deviation in the vicinity of the maximum, 
and the asymptotics near the tail. The tail is governed by high order cumulants,
whereas the center of the distribution is determined by the lowest
non-trivial cumulants---i.e., the third and the fourth cumulants (n=3 and 4).
Recent calculations in random matrix theory \cite{macedo}, based on 
a local maximum-entropy approach in the Landauer picture, found that 
for a unitary ensemble ($B \ne 0$) the third cumulant vanishes
in quasi-one dimension \cite{definition}. This result was confirmed by 
calculations in the microscopic diagrammatic theory \cite{rossum1}, which 
further added that the third cumulant or the skewness is negative in quasi-2D 
and positive in 3D.

In the experiments reported in this Letter, we find that (a)
the conductance distribution in quasi-1D metallic ($g \ge 8$) wires is
not symmetric, (b) the third cumulant is non-zero and large, and furthermore, 
(c) it is positive. The experimental artifacts as well as the effects of temperature
and dephasing are expected to go in the opposite direction---that is, they 
work to symmetrize the distribution and reinforce the gaussian. 
With the increase of temperature from 38 mK to 300 mK, the asymmetry indeed vanishes.
The observed skewness at low temperature is 
\twocolumn[\hsize\textwidth\columnwidth\hsize\csname @twocolumnfalse\endcsname
\begin{figure}
 \vbox to 5cm {\vss\hbox to 14cm
 {\hss\
   {\includegraphics{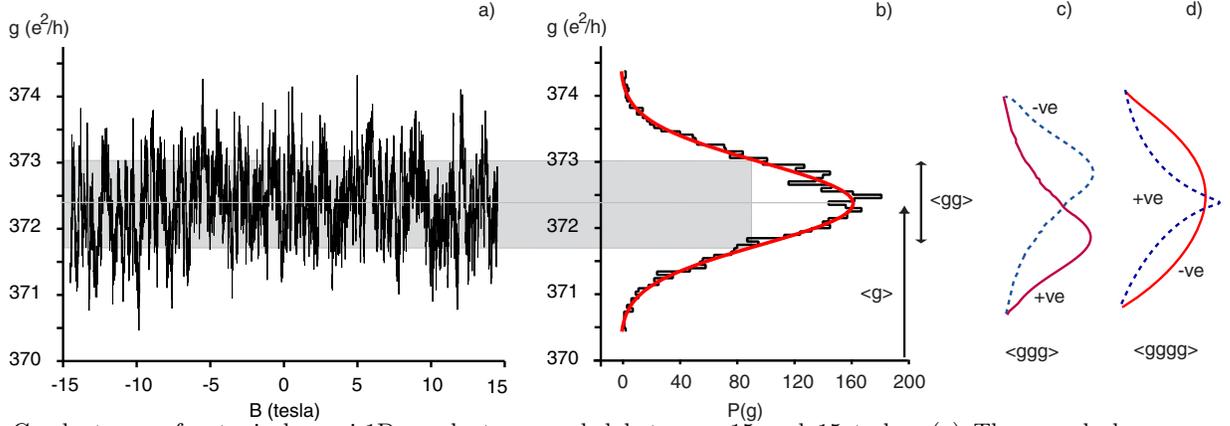}
   }
  \hss}
 }
\caption{Conductance of a typical quasi-1D conductor sampled between
-15 and 15 tesla. (a) The sample has a mean conductance $\langle g \rangle$ of
372.3. (b) The conductance distribution $P(g)$ 
constructed from (a). $P(g)$ is gaussian around the mean $\langle g \rangle$,
and hence, dominated essentially by universal conductance fluctuations 
(UCF) or the second cumulant (variance) $\langle gg \rangle$, which 
determines the full width at half maximum.
(c) A distribution whose third cumulant 
$\langle ggg \rangle$ is substantially different from that of a normal 
gaussian distribution is schematically shown.(d) The schematic representation
shows a distribution whose fourth cumulant $\langle gggg \rangle$ varies
greatly from a gaussian. 
}
\end{figure}
]
\noindent
not reminiscent of a log-normal tail expected at the onset of localization. All this suggests
the failure of perturbation theory with renormalized conductance
allowed by the scaling procedure, which is equivalent to the summation of all the 
quantum corrections. As is known, the divergent terms corresponding to 
the quantum corrections cannot be summed for $d \le 2$ in a perturbative 
framework with the un-renormalized classical conductance. In short,
a substantial deviation from the gaussian in the metallic region strongly
suggests the breakdown of one-parameter scaling procedure, since one needs
more than one parameter---the higher order cumulants, to describe the distribution. 
Its failure in $d < 2$ is serious, as this is where it is supposed to be 
robust and free from theoretical problems \cite{kumar}. However, previous 
experiments on conductance distribution have not addressed 
this issue \cite{dominique}. 

We have measured the full conductance distribution in a set of 
quasi-1D gold conductors with two high-conductance (high-g) and 
two low-conductance (low-g) samples, whose dimensions are given
in Table I. The dimensionless conductance in the samples is
 372, 473, 10.8, and 8.9 at 4 K. 
The high-g samples, 1d-A and 1d-B, are phase-coherent rings with $L < L_\phi$,
the decoherence length.
The low-g samples, 1d-C and 1d-D, are long wires; the elastic mean free path 
$l_e$ is 13.3 nm, and
11.0 nm, and diffusion constant $D$ is 6.06 $\times 10^{-3}$ m$^2$/s, 
and 5.02 $\times 10^{-3}$ m$^2$/s. From the weak localization measurements $L_\phi$ 
in 1d-C and 1d-D samples is found to be $\sim$ 4 $\mu$m at 38 mK. These samples are 
designed for two-probe measurements with no other outcoming leads to 
allow for proper comparison with random matrix theory, valid only for 
two-probe measurements. We have also measured the conductance distribution 
in other quasi-1D wires and rings in two-probe and four-probe configurations. 
Our experimental findings reported here are consistent with the data
obtained in these additional samples--that is, the high-g
samples and the low-g samples behave differently.
 
The ensemble for generating the conductance distribution is 
created by sweeping the magnetic field up to $\pm$ 15 tesla. 
In the construction and the analysis of the distribution, 
we have taken following considerations into account: 
(a) In order to obtain a large number of statistically independent 
intervals, samples were designed to have large width $w$ and $L_\phi$ 
for short correlation scale $B_c \sim {h/e \over wL_\phi}$. However,
the requirement of quasi-one dimensionality prevented us from increasing
$w$ arbitrarily. The samples studied here represent the optimized set 
of parameters under the present experimental conditions. 
(b) Data sets containing instrumental fluctuations were removed. This was 
achieved by conducting the magnetic field sweep repeatedly, and identifying
the large irreproducible fluctuations. Additionally, the positive-field ($B > 0$)
and negative-field ($B<0$) parts of the data were compared
for the $g(B)=g(-B)$ symmetry in the case of two-probe samples. In the
analysis of the data from the four-probe samples, the asymmetric part, 
which has its origin in the non-local contributions \cite{benoit}, was removed. 
(c) To check for the consistency in our analysis, cumulants were also determined
from the data binned into sizes larger than the correlation
scale $B_c \sim {h/e \over wL_\phi}$. This ensures that the binned
data points are statistically uncorrelated even in the sense of UCF--that is, each 
binned data point represents a conductance fluctuation due to an independent
set of interference paths; this procedure reduces the number of points to typically 1000,
and hence detriments the statistics by a factor of 2 (for 1d-A and 1d-B) and 3 (for 1d-C and
1d-D). (d) For the calculation of the cumulants, data points between -0.04 to +0.04 tesla,
the weak localization corrections, were removed. Thus the  
data truly represents a unitary ensemble for the purpose of comparison to 
random matrix theory. 

\twocolumn[\hsize\textwidth\columnwidth\hsize\csname @twocolumnfalse\endcsname
\begin{figure}
 \vbox to 6 cm {\vss\hbox to 14cm
 {\hss\
   {\includegraphics{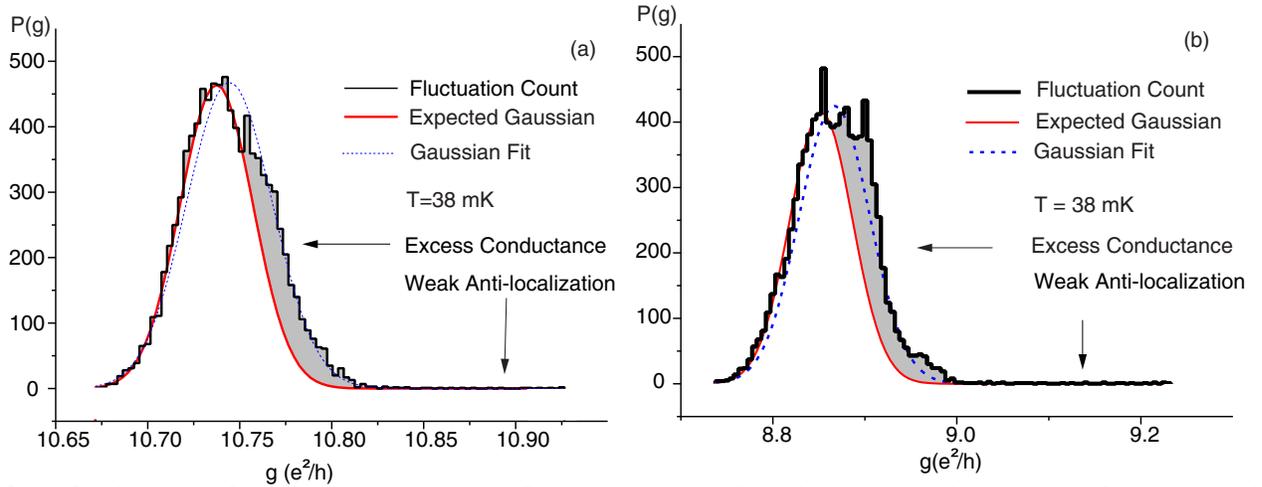}
   }
  \hss}
 }
\caption{Conductance of two representative quasi-1D Au wires at 38 mK with
a mean conductance $\langle g \rangle$ of (a) 10.75 and (b) 8.85. 
The histograms are constructed from an ensemble of $\sim$ 9400 samples.
The gaussian part of the distribution displays the UCF 
contribution. The long tail towards higher conductance represents
the weak-anti-localization contribution (not included in the evaluation of the
cumulants). Unlike in Fig.~1, the distributions
are manifestly asymmetric. The shaded areas depict the deviation from 
the expected gaussian by the higher order contributions $(2 < n \ll g_0)$; and  
the shape indicates a large third cumulant
$\langle ggg \rangle$ with a positive coefficient. The dotted curve in both
figures is the best gaussian fit to the data.}
\end{figure}
]

Fig.~1 displays a typical magnetoconductance trace for the sample 1d-A. 
We identify the first cumulant---the mean $\langle g \rangle$, and
the second cumulant--the variance $\langle gg \rangle$. 
The UCF contribution $\langle gg \rangle$ is the universal part of 
the distribution, independent of $\langle g \rangle$ or $l_e$. This
UCF part is a gaussian centered---and symmetric---around 
$\langle g \rangle$. Fig. 1(c) shows the schematics for distributions
whose third cumulant $\langle ggg \rangle$ or the asymmetric deviation 
from the gaussian is large. The fourth cumulant $\langle gggg \rangle$ or 
kurtosis, shown schematically in Fig. 1(d), determines the symmetric shape 
deviation; its positive or negative values represent deviations leading 
to sharp or flat distributions respectively.
As shown in Fig.~1(b), the distribution for the sample 1d-A
is a gaussian with undetectable deviations up to the fourth order.

Fig.~2 shows the distributions $P(g)$ for the two low-g samples at 38 mK. 
In both the samples, 1d-C and 1d-D, a strong deviation
from the gaussian is observed with the striking aspect that the asymmetry is present on
the high-g side of the mean, and the low-g part of the distribution fits
to a gaussian extremely well. The deviation is predominantly 
in the vicinity of the maximum, and is not reminiscent of a log-normal tail.

Temperature and dephasing tend to reduce the size of the fluctuations and
make the distribution gaussian. By raising the temperature to 300 mK,
we indeed find that $P(g)$ becomes a gaussian in both the samples, as shown 
in Fig.~3. This also serves as an important experimental check to ensure that the 
deviations observed at low temperature are not due to instrumental artifacts.
Quantitative estimate of the deviation is difficult, considering the
number of statistical independent data points accumulated between 
-15 and 15 tesla. N = 4800 for 1d-A and 1d-B, and N = 9391 for 1d-C and
1d-D.  The variance  $\langle gg \rangle$ in the high-g samples ($L < L_\phi$) 
is consistent with the UCF theory \cite{fukuyama}, which includes temperature through energy averaging. Data from the low-g samples agree with theory, which includes both temperature ($L_\phi > L_T$, the Thouless length) and dephasing ($L_\phi < L$) effects:
\begin{equation} 
\langle gg \rangle_{expt} \simeq {8\pi \over 3} {L_\phi L_T^2 \over L^3}.  
\end{equation}

The third cumulant, $\langle ggg \rangle$ is within
the statistical error bar for the high-g samples 1d-A and 1d-B. However,
the low-g samples have a large numerical value of $\langle ggg \rangle$,
as would be expected from Fig.~2. 
The positive value argues that the two samples are not in the
quasi-2D regime, where the expected value is both negative and small 
($-0.0020\langle g^{-1} \rangle \sim -0.0002$). Beacuse of poor statistics,
it is difficult to discern the fourth order deviation
$\langle gggg \rangle$ from the data, though it appears to be non-zero and negative.
 
It is clear that the third cumulant of conductance distribution is non-zero, 
contrary to the random matrix theory and microscopic diagrammatic theory. Both 
theories do not account for interaction. The first possibility is the dominant 
role of interaction. As seen in Fig.~2 and 3, going from 38 mK to 300 mK,
the mean conductance changes by an amount $\delta \langle g \rangle_{ee}$, larger 
than the conductance fluctuations. One would expect that interaction corrections
to the UCF part $\langle gg \rangle$ and higher order cumulants can be large 
enough to distort the gaussian. As the one-parameter scaling results in the 
gaussian distribution, assuming the Einstein relation between the conductance 
and the diffusion coefficient $\sigma = e^2 N(0) D$, the effect of interaction on the fluctuations of density of states $N(0)$ may be important.
\twocolumn[\hsize\textwidth\columnwidth\hsize\csname @twocolumnfalse\endcsname
\begin{figure}
 \vbox to 5cm {\vss\hbox to 14cm
 {\hss\
   {\includegraphics{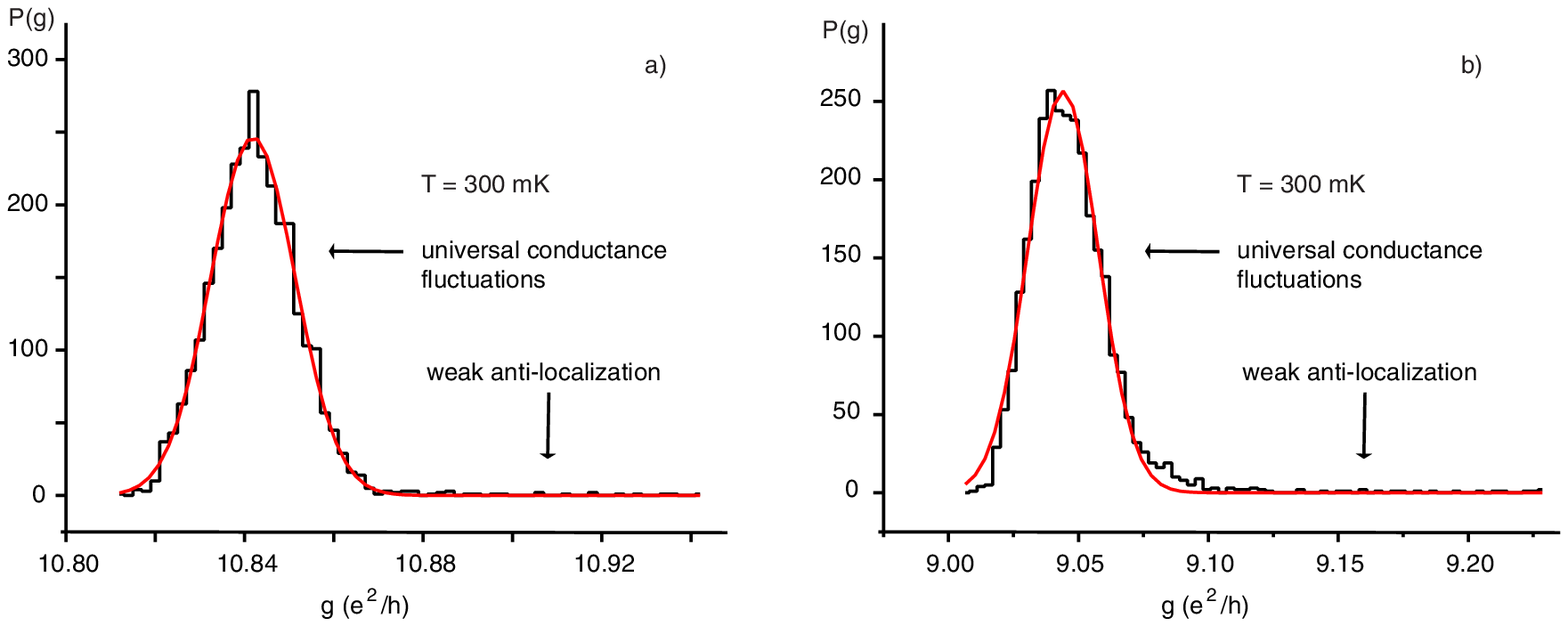}
   }
  \hss}
 }
\caption{Conductance of the same two quasi-1D Au wires (displayed in Fig.~2) 
at 300 mK with respective mean conductances of (a) 10.84 and (b) 9.05;
the shift in the mean is due to the electron-electron interaction. }
\end{figure}
]

Another possibility is that the perturbative treatment of conductance
in terms of $1/g$ as the expansion parameter is inadequate,
even though g is large enough ($g \sim 8-10$) for the perturbative expansion 
to hold. The non-zero value of $\langle ggg \rangle$,
contrary to the diagrammatic calculations, may require a new 
non-perturbative analysis.
 
In summary, we report the observation of non-gaussian conductance distribution in
low-conductance metallic quasi-1D wires at low temperature (with greater 
than 3 sigma confidence level). This is not expected in random matrix theory or 
perturbative diagrammatic theory. The necessity of additional parameters, or
cumulants, to describe the conductance possibly signals the breakdown of
the one-parameter scaling hypothesis. This research is supported by
NSF Gant No. DMR-9510416 and NSA Grant No. MDA90498C2194.

\twocolumn[\hsize\textwidth\columnwidth\hsize\csname @twocolumnfalse\endcsname
\begin{table}{Table 1. The mean $\langle g\rangle$ and the higher order moments
of conductance for two high-g and two
low-g metallic samples at 38 mK. The sampling errors in $\langle ggg \rangle$ and 
$\langle gggg \rangle$ with respect to a normal distribution 
are $\sqrt{6/N}$ and $\sqrt{96/N}$ respectively \cite{kendall}. 
Note that $\langle ggg \rangle$
is not affected by Sheperd's correction for grouping \cite{kendall}.}
\vskip 0.1in
\begin{tabular}{lccccccccc}
Sample & L ($\mu$m) & w(nm) & t(nm) & R($\Omega$)  & $\langle g\rangle$ & $\langle gg\rangle$ & $\langle ggg\rangle$ & $\langle gggg\rangle$  \\ \tableline 
1d-A  & 2.95 & 35.7   & 22     & 69     & 372.39  &  0.33     & -0.015$\pm$0.035 	& -0.29$\pm$0.14 \\ 
1d-B  & 2.95 & 26.7   & 22     & 56     & 473.39	 &  0.65     & -0.020$\pm$0.035 	& 0.17$\pm$0.14  \\
1d-C  & 20.3 & 30.0   & 18     & 2390   & 10.75	 &  0.00055  & 0.164$\pm$0.025 	& -0.27$\pm$0.10 \\  
1d-D  & 20.3 & 30.0   & 18     & 2886   & 8.85	 & 0.00171   & 0.087$\pm$0.025 	& -0.06$\pm$0.10 \\
\end{tabular}
\end{table}
]


\begin{references}
\bibitem{definition}
``Quasi-one dimension'' in the models based on the Landauer formalism implies 
that the cross-sectional dimensions are smaller than---or, comparable to---the 
mean free path for elastic scattering $l_e$ and much larger than the fermi wavelength.
\end{references}
\end{document}